# Surface and Temporal Biosignatures

Edward W. Schwieterman
*Department of Earth Sciences, University of California, Riverside, CA, USA*
*Blue Marble Space Institute of Science, Seattle, WA, USA*
*e-mail: edward.schwieterman@ucr.edu*

## Contents

**Abstract**
Recent discoveries of potentially habitable exoplanets have ignited the prospect of spectroscopic investigations of exoplanet surfaces and atmospheres for signs of life. This chapter provides an overview of potential surface and temporal exoplanet biosignatures, reviewing Earth analogues and proposed applications based on observations and models. The vegetation red-edge (VRE) remains the most well-studied surface biosignature. Extensions of the VRE, spectral 'edges' produced in part by photosynthetic or nonphotosynthetic pigments, may likewise present potential evidence of life. Polarization signatures have the capacity to discriminate between biotic and abiotic 'edge' features in the face of false positives from band-gap generating material. Temporal biosignatures — modulations in measurable quantities such as gas abundances (e.g., $CO_2$), surface features, or emission of light (e.g., fluorescence, bioluminescence) that can be directly linked to the actions of a biosphere — are in general less well studied than surface or gaseous biosignatures. However, remote observations of Earth's biosphere nonetheless provide proofs of concept for these techniques and are reviewed here. Surface and temporal biosignatures provide complementary information to gaseous biosignatures, and while likely more challenging to observe, would contribute information inaccessible from study of the time-averaged atmospheric composition alone.

# Introduction

Advances in the discovery and characterization of exoplanets in recent years have enhanced the prospects for characterizing planets within the circumstellar habitable zones of nearby stars. Providing an exoplanet meets the minimum criteria for planetary habitability (e.g., the capacity to maintain stable surface liquid water), attention will turn to the search for evidence of life. Such evidence of a living process is termed here as a "biosignature." Formally, and most generally, a 'biosignature' has been defined as an "object, substance, and/or pattern whose origin specifically requires a biological agent" (Des Marais & Walter 1999; Des Marais et al. 2008). Evidence for life on an exoplanet must be determined by remotely available information, detectable over interstellar distances. Therefore, the presence of an exoplanet biosignature can only be determined by the pattern imprinted by life upon the electromagnetic energy scattered, reflected, emitted, or transmitted from the planet to the observer. However, the distance and relative dimness of a putative habitable exoplanet will make inferences of the evidence of life quite challenging. In contrast to other fields, such as palaeontology and geochemistry, an exoplanet biosignature is unlikely to ever be completely unambiguous, so it is always best to think of an exoplanet biosignatures as a "potential biosignature."

There is currently no universally recognized categorization scheme for exoplanet biosignatures, although some attempts at this have been made (e.g., Seager et al. 2012; also see Walker et al. 2018). One suggestion is to broadly group potential biosignatures in terms of how they will manifest themselves to an observer. In this scheme three broad types of exoplanet biosignatures exist: gaseous products of life, surface signatures of living material, and temporal modulations of measurable quantities such as gas concentrations attributable to life (following Meadows 2006, 2008).

Specific gases are produced by life on Earth and their presence in an exoplanet atmosphere, inferred from the spectroscopic signatures of the gas(es), may indicate life if abiotic origins can be reasonably excluded. The most commonly referenced biosignature gases for Earth-like ($N_2$-$CO_2$-$H_2O$ dominated) atmospheres are oxygen ($O_2$), its photochemical by-product ozone ($O_3$), methane ($CH_4$), and nitrous oxide ($N_2O$), and combinations thereof (Meadows and Seager 2010). Exoplanets with atmospheres more like a younger Earth may host different types of atmospheric biosignatures, such as sulfur bearing gases or hydrocarbons (e.g., Pilcher 2003; Domagal-Goldman et al. 2011; Arney et al. 2017; Schwieterman et al. 2018). Photochemistry is also predicted to control the relative build-up of certain gases, making the detectability of biosignature gases such as $CH_4$ and $CH_3Cl$ dependent on the host star spectral type (e.g., Segura et al. 2005; Rugheimer et al. 2015). The interpretation of gaseous biosignatures is potentially fraught with 'false positives' from abiotic planetary processes that may mimic living processes (e.g., Harman et al. 2015). Accordingly, attention must be paid to the planetary environment and clues afforded by other gases in the planet's atmosphere to evaluate the origin of a putative biosignature gas (e.g., Schwieterman et al. 2016; Meadows 2017; and see "Biosignature False Positives" in this volume). The nature of gaseous biosignatures is more fully discussed in the accompanying chapter "Atmospheric Biosignatures."

A surface biosignature results from living material imprinting an inferable spectral or polarization marker on reflected, transmitted, or scattered light. A temporal biosignature is a modulation of an observable quantity that can be linked to a living process. This could be a seasonal change in strength or location of a surface signature, or it could be a modulation of a spectrally observable gas such as $CO_2$ due to global changes in the balance between photosynthesis and respiration linked to seasonality. Alternatively, the temporal signature could be produced by direct emission of light by living organisms, through fluorescence or bioluminescence. Both surface and temporal biosignatures are less well studied than gaseous biosignatures due to their more complex manifestations. For example, accurate modelling of the appearance of surface reflectance signatures requires additional assumptions about the interaction of light with living material (e.g., transmission and scattering in a cell or community architectures) in addition to modelling the radiation from the star that enters the planetary atmosphere, reaches the surface, and is reflected or scattered towards the observer. This chapter describes possible range of surface and temporal biosignatures that may manifest themselves in the spectral character of an exoplanet.

In examining potential surface and temporal biosignatures, we draw extensively on analogies to surface and temporal signatures of life observable on Earth. This is quite simply because Earth hosts the only known reservoir for life, and more importantly, it is the only body in the solar system for which evidence of life is remotely detectable. This does not mean we exclude the possibility of different types of life. However, the evidence for life must be remotely detectable and we must have the capacity to recognize it. The search for analogues to Earth life, or conceptual extensions thereof, is the central and essential conceit of exoplanet biosignature science in its current state (but see also Walker et al. 2018).

## Surface Signatures of Life

This section describes the origin and manifestation of signatures of surface life, using life on Earth as an analogy for what may be discovered on exoplanets. Both the origin of the signature and the type of life that generates it are discussed.

### *Photosynthesis*

Photosynthesis is the process by which life uses photon energy from the Sun (or, more generally, a star) in combination with gaseous or solid substrates to generate chemical energy. This energy can be used to generate new biomass or meet the other energy demands of the organism. A photosynthesizer is a life form that performs photosynthesis. A photosynthesizer is also a type of "primary producer," which means it generates the organic matter and metabolically accessible energy that other organisms depend upon. Because photosynthesizers have direct access to abundant energy from the Sun, they are the most productive form of life on Earth, far outpacing chemosynthetic metabolisms that rely on chemical energy gradients already present in the environment (Des Marais 2000). Because photosynthesis leverages abundant stellar energy, is sensible then to assume that photosynthesizers, if present on an exoplanet's surface, would be one of the most detectable forms of life as well, since they would have access to the largest energy source by several orders of magnitude (Kiang et al. 2007a).

The biochemical details of photosynthesis are quite involved, and beyond the scope of this chapter (for a detailed review see Blankenship 2002). However, when considering the potential manifestations of photosynthesis on an exoplanet, it is useful to know some simple chemical details including the essential reactants and products. While the specific form of photosynthesis may differ drastically on other worlds, the fundamental physics and chemistry will be the same. At its core, photosynthesis is a reduction-oxidation, or "redox," reaction where photons are collected to excite electrons from protein-pigment complexes. The electrons are transported to acceptor molecules along an "electron transport chain" and are replaced by reductants (electron donors) found in the environment. The availability of these reductants will determine, in part, the productivity of the photosynthesizers and, by extension, their ability to produce observable signatures. The entire light energy harvesting processes, from photon absorption to export of stable products, is accomplished by pigment-protein complexes termed photosystems. The two known photosystems are termed Photosystem I (PSI) and Photosystem II (PSII). These photosystems have peak absorption wavelengths of 680 and 700 nm, respectively.

A general empirical equation for photosynthesis is:

$$CO_2 + 2H_2A + h\nu \rightarrow (CH_2O)_{organic} + H_2O + 2A \qquad (1)$$

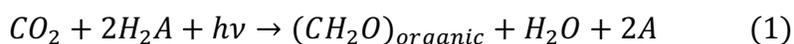

where $CO_2$ is carbon dioxide, $H_2A$ is a reducing agent (electron donor, e.g., $H_2$, $H_2S$, or $H_2O$), $h\nu$ is the photon energy needed for the reaction, and $CH_2O$ represents biomass. The use of other reductants that don't fit this equation are also possible, such as $Fe^{2+}$ and elemental sulfur (Olson 2006). If the reductant is water ($H_2O$), equation (1) becomes the equation for oxygenic photosynthesis, which generates molecular oxygen ($O_2$) as a waste product:

$$CO_2 + 2H_2O + h\nu \rightarrow (CH_2O)_{organic} + H_2O + O_2 \qquad (2)$$

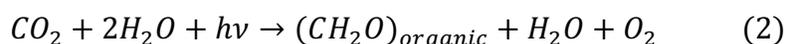

All forms of photosynthesis other than oxygenic photosynthesis are termed "anoxygenic photosynthesis", and involve electron donors other than $H_2O$ and do not produce $O_2$ as a waste product. Oxygenic photosynthesis uses both photosystems, while anoxygenic photosynthesis uses only one or the other. Physical, chemical, and molecular evidence suggests that anoxygenic photosynthesis developed first, perhaps predating oxygenic photosynthesis by up to a billion years (Buick 2008).

### *Signatures of Photosynthesis: Photosynthetic Pigments*

The absorption of light by photosystems is facilitated by photosynthetic pigments (Scheer 2006). Oxygenic photosynthesizers such as plants, algae, and cyanobacteria use chlorophyll pigments. There are several different types of chlorophyll, with chlorophyll *a* (Chl *a*) and chlorophyll *b* (Chl *b*) being the most common types in land vegetation. Both (Chl*a*) and (Chl *b*) possess absorption peaks in the red and blue, with Chl *a* peaking at 430 and 662 nm while Chl *b* peaks at 453 and 642 nm (though the location of these peaks shift depending on the cell they are contained within and the solvent used to extract them; see, e.g., Kobayashi et al. 2013). The green coloration of chlorophylls result from less efficient absorption in the green part of the visible spectrum in between the blue and red absorption peaks. Algae and cyanobacteria contain additional types of chlorophyll pigments — such as Chlorophylls c, d, and f — that absorb light at slightly different wavelengths (Kiang et al. 2007b).

Anoxygenic photosynthesizers such as purple bacteria, green sulfur bacteria, heliobacteria, and filamentous anoxygenic phototrophs (FAP) use alternative, but related, pigments called bacteriochlorophylls (Scheer 2006, Blankenship 2010). Bacteriochlorophylls (Bchls), like chlorophylls, have short and long wavelength absorption peaks, but unlike chlorophylls, the long wavelength peaks occur in the near-infrared. For example, Bchls *a* and *b* have *in vivo* (in the cell) peaks around 400 nm, 600 nm, and 800 nm with an additional peak at < 900 nm in Bchl *a* and at 1000-1040 nm for Bchl *b* (Cogdell and van Grondelle 2003; Scheer 2006). Bchl *b*'s long-wavelength peak is near one hypothetical limit for the excitation energy capable of completing electronic transitions, around ~1100 nm, although the definite limit has not yet been found (Kiang et al. 2014). In addition to the primary light capture pigments, photosynthesizers use "antennae pigments" such as types of carotenoids to capture higher energy photons and transfer them to the photosystem. Figure 1 shows the absorbance spectra of a variety of selected chlorophylls, bacteriochlorophylls, and other light-capturing pigments.

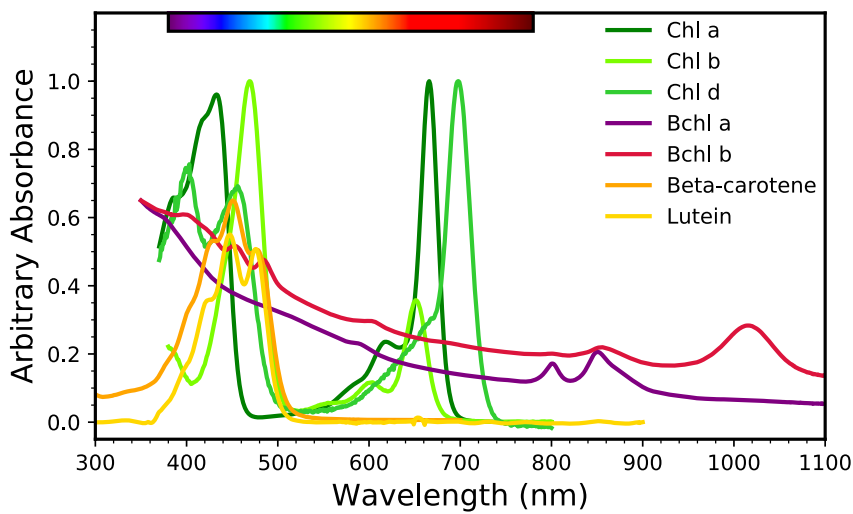

**Fig. 1** *Photosynthetic and other light-absorbing pigments.* Absorbance spectra of selected chlorophylls (Chls), bacteriochlorophylls (Bchls), and carotenoids showing the wavelengths of absorption peaks. A visible spectrum colorbar is shown at the top. The y-axis scaling is arbitrary. The Chl absorbance data are of extracted pigments in methanol solution from Chen & Blankenship (2011). The Bchl *a* and Bchl *b* data are of whole cells and originate from Cogdell & van Grondelle (2003). The beta-carotene (carotenoid) spectrum is of pigment extract in hexane from Dixon et al. (2005), while the lutein (carotenoid) is from Janik et al. (2008). Pigments dissolved in solvents have absorption peaks slightly shifted from those in cells. These data are publicly available on the Virtual Planetary Laboratory Biological Pigments Database (http://vplapps.astro.washington.edu/pigments).

*Signatures of Photosynthesis: The Vegetation Red-Edge*

The Vegetation Red-Edge (VRE) is the most well-studied surface signature of vegetation, widely suggested as a model for a planetary biosignature (Sagan et al. 1993; Arnold et al. 2002; Des Marais et al. 2002; Woolf et al. 2002; Seager et al. 2005, and many others). The VRE is the sharp increase in the reflectance of oxygenic photosynthesizers near the boundary between visible and near-infrared wavelengths (~700 nm), particularly apparent in green vascular plants (Gates et al. 1965). The VRE is produced from the contrast between chlorophyll absorption at red wavelengths (~650 – 700 nm) and the scattering properties of the cell and leaf structure at NIR wavelengths (~750-1100 nm). This contrast results from the lack of pigment absorption at NIR wavelengths and change of index of refraction between the cell walls and the intercellular space (Knipling 1970; Seager et al. 2005). Oxygenic photosynthesizers, including moss, lichen, plants, seagrass, and algae have VRE break wavelengths clustered between 0.69-0.73 μm (Kiang et al. 2007b). The VRE is a significantly strongly feature than the weaker chlorophyll "green bump." This strength is demonstrated in Figure 2, which shows the reflectance spectra of vegetation and a selection of other oxygenic photosynthesizers (lichen, algae, and a bacterial mat), illustrating the red-edge effect and contrasting the features with other common surface types (soil, snow, and seawater).

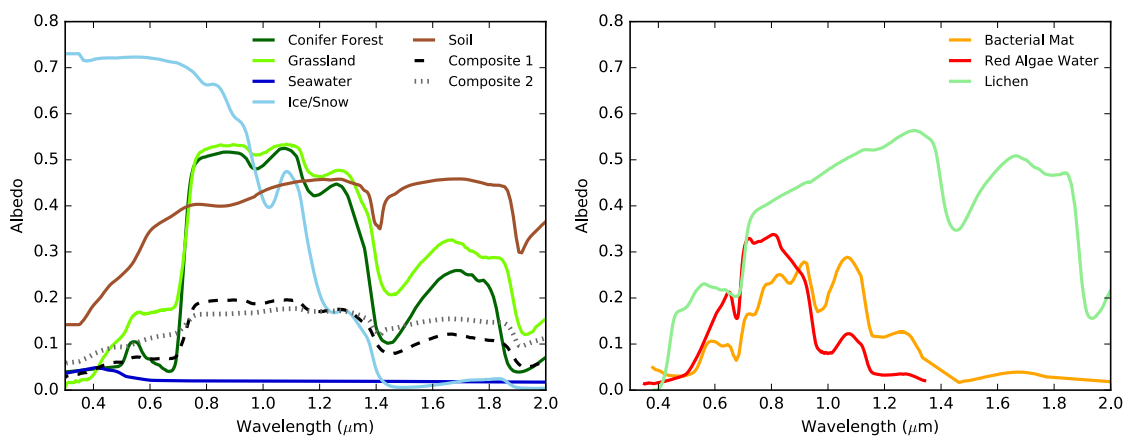

**Fig. 2** *The "Red-Edge" biosignature near 0.7 μm.* (*Left*) Albedo spectra of red-edge producing vegetation (forest, grassland) and other common surface types (seawater, ice, and soil). The composite 1 albedo consists of 66% seawater, 14% grassland and brush, 4% conifer forest, 5% bare soil, and 11% snow and ice, consistent with the an equatorial view of Earth during spring equinox (Robinson et al. 2011). The spectrum labelled composite 2 is the same as composite 1 except with the weighting for vegetation replaced by bare soil. (*Right*) Albedo spectra of other red-edge producing oxygenic photosynthesizers. The conifer forest spectrum is from the ASTER spectral library (Baldridge et al. 2009), while all others are from the USGS spectral library (Clark et al. 2007).

The two modes by which the VRE has been quantified in Earth's spectrum are measurements of the Earthshine reflected by the Moon (Arnold et al. 2002; Woolf et al. 2002; Seager et al. 2005; Montanes-Rodriguez et al. 2005, 2006; Hamdani et al. 2006; Turnbull et al. 2006; Arnold 2008; Sterzik et al. 2012) and by direct observations by interplanetary spacecraft (Sagan et al. 1993; Livengood et al. 2011). The study of Earth's spectrum by the Galileo probe flyby described in Sagan et al. (1993) provides a strong detection of the VRE, but with spatially resolved data at 1-2 km resolution. The increase in Earth's disk-averaged NIR spectral albedo attributable to the red-edge has been estimated from Earthshine measurements and spectral models to be between ~2-10% (Arnold 2008) with significant variability caused by differences in cloud cover, vegetation covering fraction, and Earth-Moon viewing geometry. Chlorophyll in the ocean produces too weak of a signature to be detectable in the disk-average. Additional complications result when considering minerals with upward sloping features that can mimic the red-edge effect (Tinetti et al. 2006a). Figure 3 shows VPL model spectra of Earth at full phase with ocean-dominated and land-dominated views, illustrating one of the most favourable cases for the detection of the red-edge. The difference in brightness between these views is ~20% just beyond 0.7 μm, though much of this difference is due to albedo differences between ocean and bare land (see Figure 1, left panel). This model includes self-consistent cloud cover, molecular absorption, Rayleigh scattering, the modern Earth's surface configuration, and the surface albedos given in Figure 2.

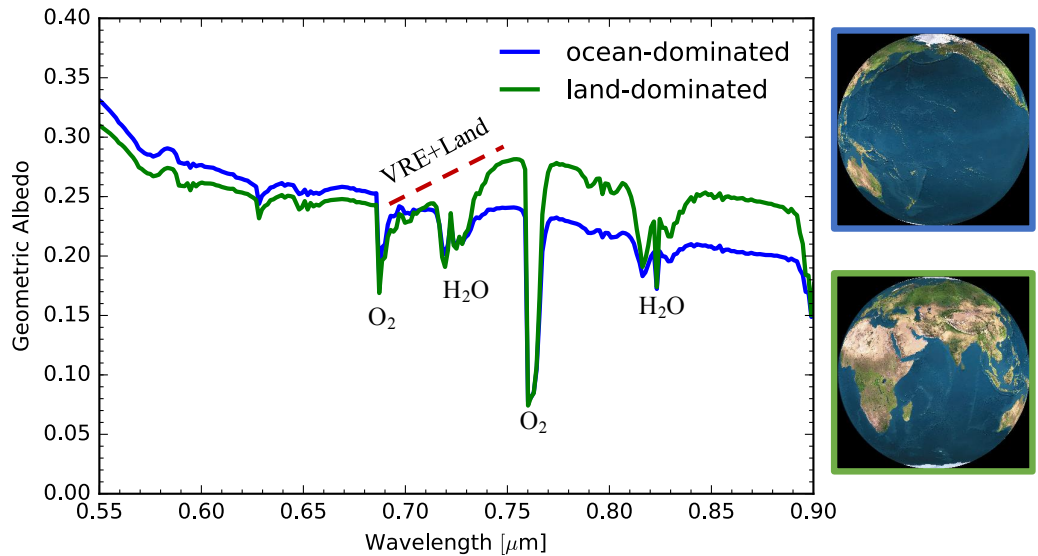

**Fig. 3** The *red-edge in synthetic disk-averaged Earth model spectra.* Synthetic disk-averaged spectra of Earth at ocean (blue) and land-dominated (green) views (consistent with Lunar viewing angles) produced using the Virtual Planetary Laboratory's 3D spectral Earth model described in Robinson et al. (2011). The increase in albedo between 0.7 and 0.75 μm is partially caused by the VRE in addition to the brightness of soil at these wavelengths (see composite surface albedos in Figure 1, left panel). Note that a water band at ~0.72 μm partially confounds the VRE signal. The images to the right show the distribution of land and ocean at the ocean and land-dominated views and were created using the Earth and Moon Viewer (http://www.fourmilab.ch/earthview).

Generally, a ~1% photometric sensitivity has been suggested as necessary to detect a VRE analogue (Arnold 2008). This is consistent with the work by Brandt and Spiegel (2014), who find that detecting the VRE would require a signal-to-noise ratio of ≥ 100 assuming 10% vegetation coverage and 50% cloud cover at a spectral resolving power ($\lambda/\Delta\lambda$) of R=20 (and an SNR ≥ 40 at R=150). Brandt and Spiegel (2014) further find that the SNR to detect the VRE is approximately six times that required to detect the $O_2$-A band.

The strong spectral break of the VRE has been leveraged to study vegetation extent and health by the Earth observing community. The normalized difference vegetation index (NDVI) was developed to identify and characterize vegetated areas on Earth's surface (Tucker 1979; Tucker et al. 2005). The NDVI is calculated as

$$NDVI = \frac{NIR-VIS}{NIR+VIS} \qquad (3)$$

where NIR is the recovered albedo in the near-infrared and VIS is the albedo recovered in the visible band (Tucker 1979). The NDVI occupies the range -1 to 1 with larger values suggesting vegetated areas (due to increased reflection in the NIR vs. visible band for plants, i.e. capturing the "red edge" effect). Non-vegetated soils tend to have small positive NDVI values while snow and clouds have negative ones. Figure 4 shows a global map of the NDVI (corrected for atmospheric effects) at two times spaced six months apart.

The utility of broadband indices like the NDVI for biosignature analysis on the disk-averaged spectra of exoplanets is likely limited, since they are less able to distinguish between sharp increases like the red-edge and the gentle slope increases of minerals. Many mineral surfaces slope up from the visible to the red, and thus could create a broadband signal like that of vegetation. For example, Livengood et al. (2011) found the NDVI of the lunar surface more closely matched the NDVI of vegetated areas on Earth than the NDVI of vegetated areas matched the disk-averaged observations of the whole Earth. However, the red-edge could contribute to "first-pass" identification of Earth-like worlds when coupled with additional broadband data, and is less expensive to obtain than high resolution spectra. Crow et al. (2011) found that the Earth's broadband colour is distinguishable from other planets in the solar system because Earth scatters in the blue while reflecting the red, partly due to the VRE (but also the red-slope of soil and exposed rock). Krissansen-Totton et al. (2016) further found that the consequent broadband "U-shape" of Earth's spectrum — with contributions from Rayleigh scattering, the "Chappuis" ozone band from 0.5-0.7 μm, and the combination of the spectrum of the VRE and land surfaces — could uniquely identify Earth from a large universe of hypothetical uninhabited planets with spectrophotometry. Time-resolved, wavelength-dependent photometry coupled with mapping analysis may allow us to look for vegetated areas with upcoming direct imaging telescopes such as LUVOIR and HabEx (Cowan et al. 2009; Fujii et al. 2010, 2017, 2018). Additionally, far-future (beyond 2030s) concepts like the Exo-Earth mapper (Kouveliotou et al. 2014), a proposed space-based visible interferometer, would have the capability to image an exoplanet with spatial resolving power and would potentially have the capacity to identify vegetated areas in a similar matter to the Sagan et al. (1993) study of Earth by the Galileo space probe.

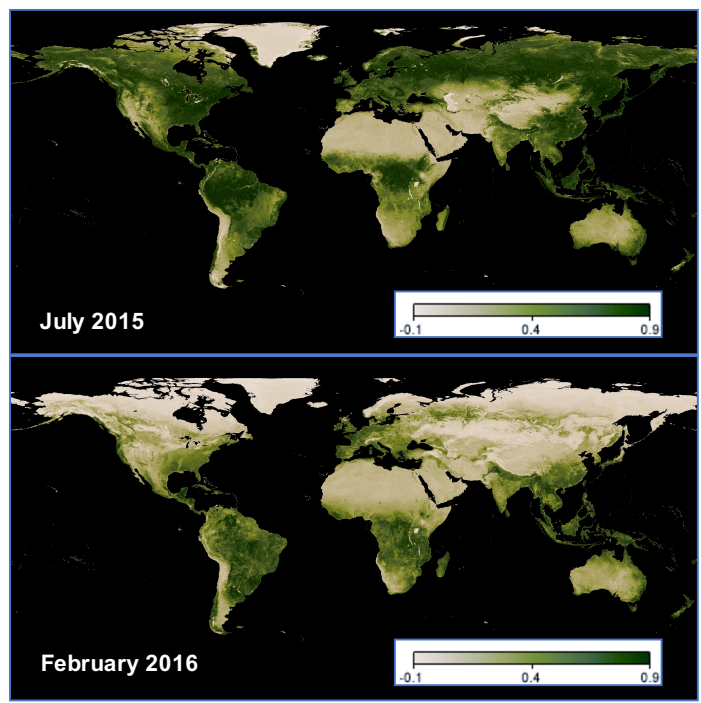

**Fig. 4** NDVI *maps*. Vegetation maps created by measuring the Normalized Vegetation Difference Index (NDVI) using the MODIS instrument aboard NASA's Terra satellite. The colour bar inset ranges from dark green (significant leaf growth) to tan (little to no leaf growth). Black areas represent no data (oceans). The NDVI is a proxy for the "red edge" spectral feature in plants. Significant differences in vegetation extent are observed between northern hemisphere summer (top; July 2015) and winter (bottom, February 2016). Credit: Images by Reto Stockli, NASA's Earth Observatory Group, using data provided by the MODIS Land Science Team (http://neo.sci.gsfc.nasa.gov/).

*Signatures of Photosynthesis Through Time*

The strength and characteristics of the VRE has likely varied greatly through geologic time. Arnold et al. (2009) modelled the VRE during the last ice age (6-21 Kya) and found the red-edge signature would be overall like that of the modern Earth, though potentially stronger or weaker depending on planet-observer geometry (e.g., weaker when viewing more extensive glaciers at the poles). Land plants have existed for only ~400 My of Earth's history (Kenrick and Crane 1997), though evidence for (anoxygenic) photosynthetic life extends to 3.5 Ga or earlier (Buick 2008). For the intervening period bacterial mats or stromatolites containing photosynthetic bacteria would have produced the most significant surface biosignatures, though their past extent and potential detectability is relatively unconstrained. Sanromá et al. (2013) found that if bacterial mats containing oxygenic cyanobacteria constituted 50% land cover or more on the early Earth, they could have been identified through spectrophotometry and distinguished from vegetation because they are dimmer at wavelengths greater than ~1.3 μm (e.g., see Figure 2). Similarly, Sanromá et al. (2014) found that bacterial mats composed of anoxygenic photosynthesizers such as "purple bacteria" will have NIR 'edges' longward of bacteriochlorophyll absorption (a ~1 µm edge) , producing a red shifted VRE effect, although significant coverage would still be required for detectability.

### *Signatures of Photosynthesis Around Other Stars*

The spectral energy distribution (SED) is shifted to shorter and longer wavelengths depending on the photospheric temperature of the host star. Kiang et al. (2007a,b) speculate that this shift in the spectrum of the star may drive the evolution of pigment absorption to the most efficient wavelength, producing 'edge' wavelengths predictable based on the stellar temperature of the host star. They note that this expected shift would be proportional to the change in the surface photon irradiance rather than energy irradiance, because photosynthesis is a quantum process requiring one or two photons (for anoxygenic and oxygenic photosynthesis, respectively) to complete (hypothetical three-photon processes have also been proposed, e.g., Wolstencroft and Raven 2002). In this scenario, a biosphere on a planet orbiting a lower temperature star (e.g., an M-dwarf) would have a NIR-shifted red-edge, while one orbiting a higher temperature star (e.g., an F-dwarf) would have a blue-shifted edge (Kiang et al. 2007a). Indeed, Tinetti et al. (2006b) find that a NIR-shifted red-edge would potentially be more detectable than the terrestrial VRE because the 'edge' would be shifted away from confounding water vapour absorption (e.g., see Fig. 3). However, the evolutionary drive to maximize photon capture may not necessarily cause a commensurate shift of 'edge' wavelengths with the spectrum of the host star. Photosynthesizers (at least on Earth) are not typically photon limited at the surface and are instead limited by the availability of water (e.g., in deserts or arid regions) or nutrients such as fixed nitrogen, dissolved iron, or phosphorous (Moore et al. 2013; Bristow et al. 2017).

### *Other Phototrophic Pigments*

Other types of pigments may produce observable signatures on exoplanets. Bacteriorhodopsin, a phototrophic pigment produced by haloarchaea, can generate catabolic energy for a cell but does not facilitate the generation of new organic matter. Its absorption spectrum is complementary to that of chlorophyll, peaking in the green (~570 nm). It has therefore been proposed that bacteriorhodopsin or a similar pigment might fuel photosynthesis elsewhere in the universe and serve as a remotely detectable biosignature (DasSarma 2006).

### *Alternative Reflectance Biosignatures*

Pigmentation has evolved for multiple purposes to adapt organisms to their environment in a variety of ways decoupled from metabolic energy capture. At times their coloration is incidental to their function. These functions of pigments, other than photosynthesis, include UV screening, reduction of free radicals, protection against extremes in temperature, regulation of growth, signalling to other organisms, and bioluminescence (see review in Schwieterman et al. 2015). For example, the carotenoid pigment bacterioruberin has been shown to assist the halophilic archaea *halobacterium salinarim* in resisting DNA damage from ionizing UV radiation (Shahmohammadi et al. 1998). Figure 5 shows the reflectance spectra of a collection of non-photosynthetic, phototrophic, and anoxygenic photosynthesizing microorganisms from Schwieterman et al. (2015). Note the diversity in edge wavelengths caused by absorption at different wavelengths and higher reflectance where pigments are not absorbing.

Hegde et al. (2015) measured the reflectance spectra from 0.35 to 2.5 μm of pure cultures of 137 microorganisms with a variety of pigments suited to different functions, generating a publically available spectral database ([biosignatures.astro.cornell.edu](biosignatures.astro.cornell.edu)). In general, these authors found that the strongest features (e.g., 'edges') in the ensemble of microorganisms were present in the visible to NIR (> 1.0 μm) range due to pigment absorption. Further into the NIR (1.0-2.5 μm), bands of hydration present at 0.95, 1.15, 1.45, and 1.92 μm appeared consistently for all samples. However, these would be problematic to observe through an atmosphere with water vapour, which absorbs near the same wavelengths.

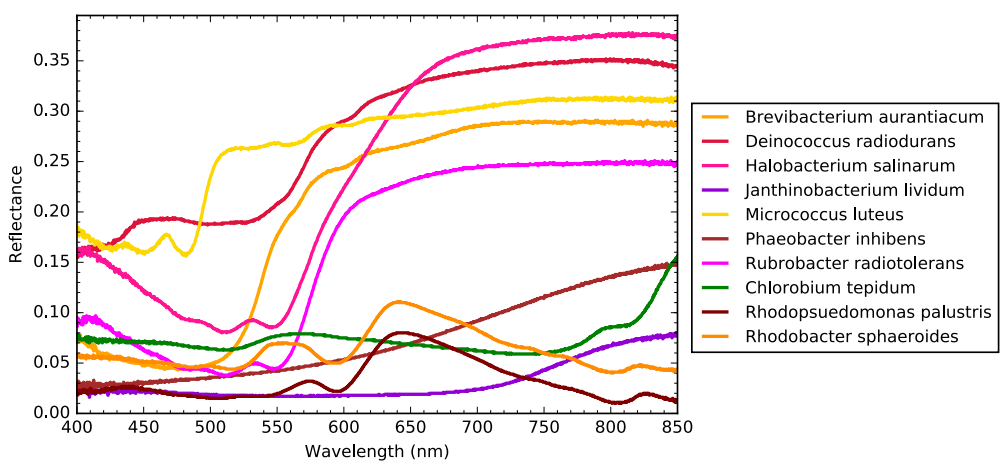

**Fig. 5** *Alternative surface reflectance biosignatures.* Reflectance spectra of a collection of non-photosynthetic, phototrophic, and anoxygenic photosynthesizing microorganisms from Schwieterman et al. (2015).

In evaluating alternative signatures to the VRE, it is instructive to consider environments on Earth where non-photosynthetic or alternative phototrophic pigments dominate the spectral reflectance rather than conventional photosynthetic pigments. An archetype for this scenario is afforded by colourful halophile-dominated salt ponds (DasSarma 2006; Schwieterman et al. 2015). The visible coloration of this environment is dominated by halophilic archaea that possess both carotenoid pigments like bacterioruberin and the phototrophic pigment bacteriorhodopsin (Oren and Dubinsky 1994). Unlike the red-edge, this alternative surface biosignature consists of a significant reflectance increase between 0.55 and 0.68 μm with a significant decline in albedo at longer wavelengths from overlying water absorption. Figure 6 shows the spectrum of a planet dominated by a forest surface and a halophile-dominated salt pond, demonstrating these contrasting surface biosignatures are potentially observable when considering the spectral effects of the atmosphere. Schwieterman et al. (2015) find that a "halophile" type biosignature could produce an albedo signatures as high as ~13% at 0.68 μm assuming surface coverage consistent with Earth's ocean fraction (~70%) and 50% cloud cover.

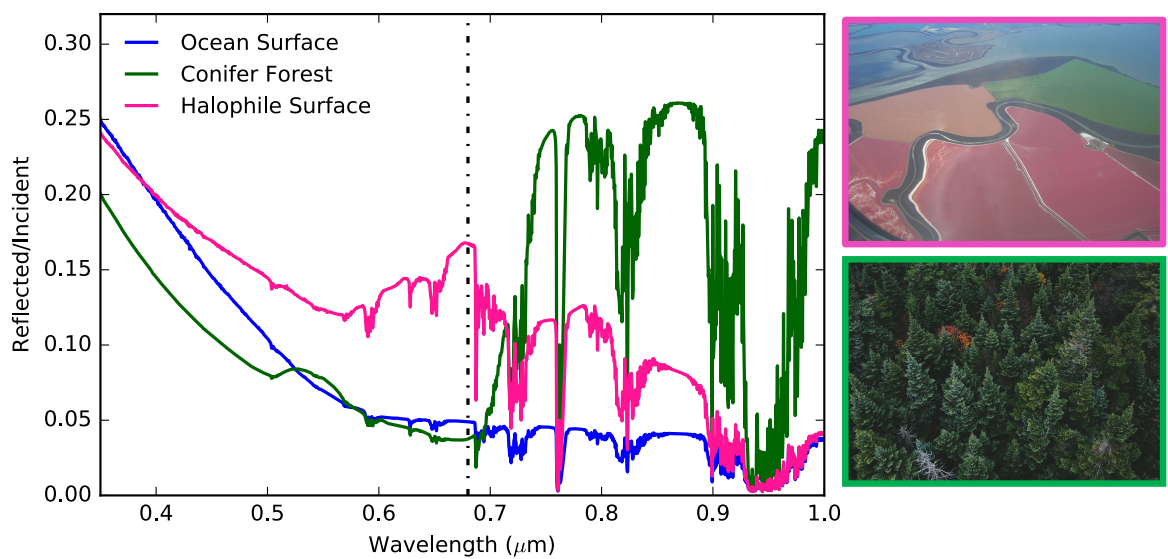

**Fig. 6** *Contrasting surface biosignatures of vegetation and halophiles.* (*Left*) Reflected light spectra of Earth-like planets dominated by ocean, forest, and pigmented halophile surfaces (after Figure 7 in Schwieterman et al. 2015). The dot-dashed line shows the characteristic halophile pond reflectance peak at 0.68 μm. The VRE from the forest is seen from 0.7-0.75 μm. The surface albedo for the forest was derived from the ASTER spectral library (Baldridge et al. 2009), and the halophile surface from a spectrum of the salt ponds in San Francisco Bay (Dalton et al. 2009). (Right) Images of the halophile-dominated salt ponds at San Francisco Bay (top; Wikipedia Commons) and a conifer forest (bottom; Pixabay).

*"False Positive" Reflectance Signatures*

The usefulness of the VRE or other similar 'edges' as biosignatures is dependent upon the extent to which they uniquely fingerprint the presence of life when compared to other surface types possible on planets with no life. The widespread use of the VRE and related NDVI index for Earth-observing applications shows this holds in practice for spatially resolved observations. However, the sharp edge features produced by electronic transitions in mineral semi-conductors have been proposed as potential false positives for red-edge analogue biosignatures (e.g., Seager et al. 2005). For example, sulfur and cinnabar have sharp 'edge-like' transitions at 0.45 μm and 0.6 μm, respectively (see Figure 7). Jupiter's moon Io has a surface that includes a substantial abundance of sulfur compounds in addition to other materials, producing a steep slope between 0.4 and 0.5 μm in the disk-average. The spectrum of Io demonstrates that these mineral surfaces may produce this reflectance effect over the surface of a planetary body (in this case a moon). This suggests that 'edge' features need to be taken in context with other planetary observables such as the composition of the atmosphere and detection of a liquid ocean through ocean glint (e.g., Robinson et al. 2010).

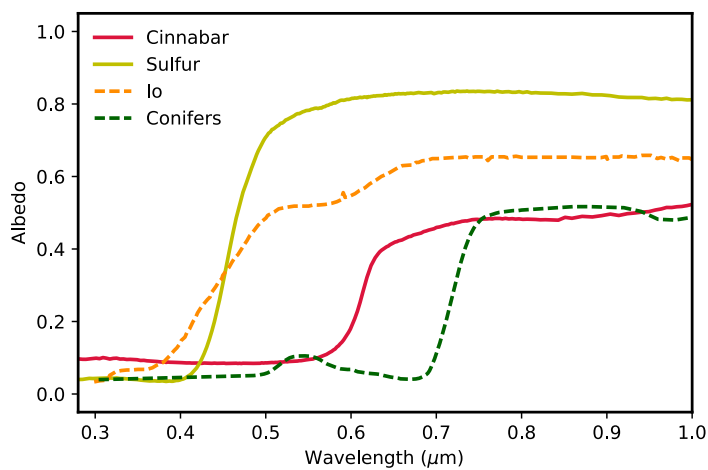

**Fig. 7** *Potential false positive 'edge' features.* Reflectance spectra of cinnabar and sulfur from the USGS spectral library (Clark et al. 2007), a reflectance spectrum of Jupiter's moon Io from Karkoschka (1994), and a conifer forest for comparison (Baldridge et al. 2009; Figure 2).

*Polarization Biosignatures*

An additional dimension may be provided in the characterization of surface biosignatures by measurements of linear and circular polarizations. In general, polarization may be a useful metric to identifying an Earth-like atmosphere, which is scattering but transparent (e.g., Stam 2008, Takahashi et al. 2013). While beyond the scope of this chapter, reviews of the radiative transfer of polarized scattering in exoplanet atmospheres can be found in several papers (e.g., Seager et al. 2000; Bailey 2007; Zugger et al. 2010; Madhusudhan and Burrows 2012).

Of particular interest for exoplanet biosignature science is linear or circular polarization imparted by photosynthetic pigments concurrently with reflectance edges. Berdyugina et al. (2016) measured the degree of polarization for a variety of leaves and flower petals, finding that the degree of polarization is maximized at wavelengths where pigment (e.g., chlorophyll, carotenoid) absorption is greatest. In other words, in vegetation polarization was high for visible (absorbing) wavelengths, and low at the NIR plateau of the red-edge reflectance increase. The degree of polarization at these wavelengths ranged from 10-90%. This trend contrasted significantly with linear polarization measurements of abiotic materials (sand, rock), which had weak wavelength dependence (Berdyugina et al. 2016). These authors additionally find strong polarization effects for modelled planetary spectra that include 80-100% coverage by pigmented organisms, though this is likely an unrealistic coverage extent. From polarization spectra of Earthshine, Sterzik et al. (2012) find that a covering fraction of 10-15% cloud-free vegetation best explains the polarization signature detected near the red-edge wavelengths and suggest that a minimum ~10% covering fraction is required to confidently detect this signature.

Circular polarization has been proposed as a unique fingerprint of homochirality (Sparks et al. 2009a, b; Patty et al. 2017). The circular polarization of vegetation is directly related to its absorption and reflection spectrum while abiotic surfaces have no correlation (Sparks et al. 2009b). The potential disadvantage to circular polarization is its relatively small polarization degree (~0.005%) and that circular polarization in remote or Earthshine observations of Earth's disk-averaged spectrum cannot yet be definitively linked to surface vegetation.

## Temporal Signatures of Life

Temporal biosignatures are time-dependent oscillations in measurable quantities, such as gas concentrations or planetary albedo, that are indicative of biological activity (e.g., Meadows 2006, 2008). This time-dependency may be seasonal, diurnal, or perhaps even stochastic, but must present an observable that can be directly linked to a consequent response of the biosphere. Temporal biosignatures are in general less well-studied than gaseous or static surface biosignatures. This is partially because a substantial number of additional variables must be considered, such as the surface (continental) asymmetry of the planet, its axial tilt and orbital eccentricity, and the putative response of the living processes that are sensitive to these factors. However, we can look to temporal variations produced by living process on Earth as proofs of concept.

### *Modulation of Atmospheric Gases*

On Earth, seasonal periodicities are observed in gases linked to the biosphere through production and consumption fluxes, which include $CO_2$, $O_2$, and $CH_4$ (e.g., see Figure 8). The oscillation of $CO_2$ is due primarily to the seasonal growth and decay of land vegetation, declining in the spring and summer and rising in the fall and winter (Hall et al. 1975). The seasonal variation in $O_2$ is stoichiometrically linked with $CO_2$ consumption by photosynthesis and production by respiration and decay of organic matter (i.e., $CO_2 + H_2O \Leftrightarrow CH_2O + O_2$), mirroring $CO_2$ variations in phase. However, because $CO_2$ has a greater solubility in ocean water, the absolute variability of $O_2$ is greater than that of $CO_2$ (Keeling and Shertz 1992), although this is a very small percentage of the total abundance of $O_2$ in the modern atmosphere. The seasonal variation in $CH_4$ is more complex with a deep minimum in northern summer, a smaller minimum in winter, and maximum concentrations in the late fall and early spring (Rasmussen and Khalil 1981). While the non-anthropogenic flux of $CH_4$ is dominated by methanogenic microorganisms in terrestrial wetlands (Cicerone and Oremland 1988), its temporal variability is only partly determined by seasonal changes in production. Rather, the annual cycle of $CH_4$ is most dominantly controlled by interactions with hydroxyl (OH) ions (i.e., $OH + CH_4 \rightarrow CH_3 + H_2O$), which are sourced primarily from tropospheric $H_2O$ (through $O(^1D) + H_2O \rightarrow 2OH$) and therefore closely track seasonal changes in surface temperatures (Khalil and Rasmussen 1983).

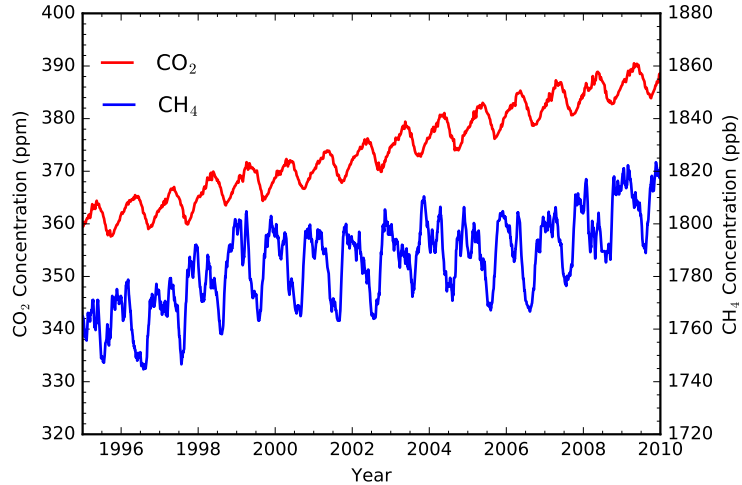

**Fig. 8** *Temporal gas variations as a biosignature.* Measurements of $CO_2$ and $CH_4$ concentrations from the NOAA at Mauna Loa, Hawaii, USA (19.5° N) from 1995-2010 (Thoning et al. 2017; Dlugokencky et al. 2017). The yearly oscillations in both gases is partly a seasonal response to changes in the biosphere's productivity in the northern hemisphere. The secular increase in the concentration of these gases is due to anthropogenic emissions. Data are sourced from the United States National Oceanic and Atmospheric Administration's Earth System Research Laboratory (https://www.esrl.noaa.gov/).

The amplitudes of the seasonal oscillations in biologically moderated gases are hemisphere- and latitude-dependent. Since the $CO_2$ oscillation is primarily driven by the terrestrial biosphere, the amplitude of seasonal changes is significantly greater in the northern hemisphere where the fractional land cover is larger (Keeling et al. 1996). The amplitude of this variation is a function of latitude, ranging from 15-20 ppm in far northern hemisphere to ~3 ppm near the equator (Keeling et al. 1996). In contrast, the seasonal variation in $O_2$, primarily driven by the ocean biosphere, is comparable between hemispheres (Keeling and Shertz 1992), though like for $CO_2$, the variations in $O_2$ are more muted for latitudes near the equator where temperature variations are smaller (Keeling et al. 1998). These hemispherical and latitudinal dependencies suggest that the detectability of seasonal cycles on exoplanets will be highly constrained by viewing geometry, emphasizing the importance of measuring parameters such as inclination and obliquity (e.g., Schwartz et al. 2016).

Detecting seasonal variations of these gases in an exoplanet atmosphere may be quite challenging. Consider attempting to measure an analogue to modern Earth's seasonal $CO_2$ cycle. Since the observation of an exoplanet will likely be a disk-average, the integrated peak-to-trough variability of $CO_2$ in the northern hemisphere is appropriate to consider, which is about 6.5 ppm or ~2% (Zhao and Zeng 2014). Per an approximation of Beer's Law at small optical depths ($\tau \ll 1$), the maximum change in $CO_2$ absorption is then ~2%, but it is likely to be smaller than that because the strongest $CO_2$ bands in Earth's spectrum (e.g., at 15 μm) are already saturated at the band cores (Des Marais et al. 2002; Robinson et al. 2011). To confidently detect a 2% variation at a 3σ significance would require a signal-to-noise (S/N) ratio of at least 150. $CH_4$ seasonal variations are comparable to those of $CO_2$ at 1-2% (Rasmussen and Khalil 1981; Khalil and Rasmussen 1983), are only partially biogenic, and would likewise be difficult to observe. This problem is even more acute for seasonal variations in $O_2$, which is dependent on latitude but varies between ~20-100 ppm against a 21% background abundance, or a ~0.01-0.05% variation (Keeling and Shertz 1992; Keeling et al. 1998; Manning et al. 2003). It is likely that seasonal variations of the above gases comparable to those of the modern Earth would be undetectably small given the capabilities of next generation space and ground-based observatories. However, planets with greater obliquity or eccentricity than Earth would have exaggerated, and perhaps more detectable, seasonal signatures. Prospects for detecting these variations would be further enhanced for cases with greater proportional changes in gas concentrations at background abundances that are spectrally detectable, but not saturated, at the band centre. A possible example of this situation may be found in the Hartley-Huggins UV $O_3$ band at $O_2$ concentrations consistent with the Earth's during the mid-Proterozoic Eon, or < 0.1% of modern oxygen levels (Reinhard et al. 2017; also see "Atmospheric Evolution of a Habitable Planet"). However, contextual information would be required to rule out large changes in mixing ratios resulting from abiotic factors, like the seasonal sublimation of $CO_2$ from the Martian polar caps (Wood and Paige 1992). Future work is required to elucidate the range of scenarios for which biological activity could be measurably inferred from the modulation of atmospheric gases on exoplanets.

*Modulation of Surface Reflectance*

Aside from gas concentrations, another possibility is that the surface signatures of life will be temporally variable. Because red-edge like features have potential false positives (e.g., Figure 7), the case for the biogenicity of an analogue "red-edge" signature would be enhanced by temporal variations in phase with observed shifts in insolation, water vapour content, or other observable parameters indicative of seasonal change. For example, the NDVI signal of dead or dying vegetation is weaker than that of living vegetation because lack of chlorophyll pigments makes leaves more reflective in the visible while desiccation renders them less reflective in the near-infrared (Tucker 1979). (Figure 4 illustrates the global spatially resolved changes in the NDVI six months apart.) Assuming views where one hemisphere dominates, or cases in which one hemisphere contains substantially more land vegetation (as is the case for Earth), this seasonality could manifest itself in the disk-average (Seager et al. 2005; Arnold et al. 2009).

Vegetation's spectral character alters in other ways as a response to seasonal changes. Specifically, the orange pigmentation of autumn leaves is produced when carotenoid pigments are uncovered by the degradation of chlorophyll, while red coloration originates from anthocyanin, produced by autumn foliage in response to drops in temperature, perhaps to provide photoprotection (Archetti et al. 2009). Bacterial mats likewise undergo seasonal changes in coloration (Nicholson et al. 1987). However, these other changes would likely be less observable than counterparts to the red-edge and would be more comparable to the much weaker "green bump," requiring substantially higher signal-to-noise ratios to observe.

Seasonal modulations in surface polarization signatures would be another possible temporal biosignature. The circular polarization signature of vegetation is dependent on the health of the organism (Patty et al. 2017) and changes as seasonal conditions cause the death and shedding of leaves. Temporal changes in surface polarization signatures could help distinguish living matter from seasonal alterations in non-living material. For example, a potential seasonal red-edge "false positive" would be melting of snow above a red-edge producing mineral like sulfur or cinnabar (e.g., Figure 7), but these nonbiotic surfaces would not show the same changing circular polarization signature as living material (Sparks et al. 2009a; Patty et al. 2017). Measuring seasonal changes in linear or circular polarization poses similar, but exacerbated obstacles compared to using polarization as a biosignature in general. Detectability threshold must reach the magnitude of the change in the polarization signature, which will be equal to, or more likely smaller than, the magnitude of the maximum polarization.

It is also important to separate seasonal temporality of the underlying biosignature with temporal fluctuations due to the rotational phase or changing cloud cover fraction observed (e.g., Ford et al. 2001). Measurements of variations in the underlying surface biosignature would require comparisons between equivalent rotational views, correction for phase effects, and an estimate of the systematic change in cloud cover, if any. In principle, these parameters can be recovered by multi-band photometry or spectroscopy at several phases (e.g., Fujii et al. 2011; Cowan & Strait 2013). It is therefore a necessity that several orbital cycles are observed to confidently detect a temporal biosignature. Such observations would be more easily facilitated by observations of a habitable planet orbiting a late type star where orbital periods are substantially shorter (Shields et al. 2016). However, inner-working angle constraints may restrict this possibility for all but the nearest targets in the stellar neighbourhood (Robinson et al. 2016).

*Biological Fluorescence and Bioluminescence*

The active emission of light by vegetation or microorganisms is yet another potential temporal surface biosignature. Fluorescence is the reprocessing of absorbed photons into less energetic emitted photons. Chlorophyll fluorescence is a signature of photosynthesis on Earth (Papageorgiou 2007), evolved to reduce physiochemical stress in high light environments, so conceivably photosynthetic pigments originating elsewhere may likewise possess similar characteristics. Chl *a* has a fluorescence spectrum characterized by broad emission between 640 and 800 nm with peaks at 685 and 740 nm (Meroni et al. 2009; Figure 9). The photons emitted at these wavelengths are reprocessed from chlorophyll absorption at shorter wavelengths (e.g., see Figure 4). Chlorophyll fluorescence has been observed by Earth observing satellites, and can be used to map ocean algae and plant health from space and determine seasonal variability (Joiner et al. 2011; Sun et al. 2017). The radiance from a fluorescent surface (*L*) can be described by the following equation (Meroni et al. 2009), assuming the Lambertian assumption applies:

$$L(\lambda) = \frac{r(\lambda)S(\lambda)}{\pi} + F(\lambda) \quad (4)$$

where *r* is the reflectance without the fluorescence contribution, *S* is the incident irradiance from the star, and *F* is the fluorescence component. For vegetated surfaces on Earth, the contribution from *F* is about 1-5% of *L* (Meroni et al. 2009; Joiner et al. 2011). Since fluorescence requires continuous UV or violet-blue visible light to stimulate emission, it can only be observed on day-lit portions of the planetary disk, which is problematic for uniquely identifying the fluorescence against the much brighter reflected light from the star. The signal then would also vary based on cloud cover extent. In Earth observing observations, one strategy is to use high resolution spectroscopy to search for the "filling in" effect of fluorescent flux on reflected solar lines or atmospheric absorption lines (Meroni et al. 2009; Joiner et al. 2011). Commonly chosen lines include the solar H$\alpha$ line at 656.3 nm, the solar potassium (K) line near 770 nm, and the atmospheric $O_2$-A and $O_2$-B bands near 761 and 687 nm, respectively (Meroni et al. 2009; Joiner et al. 2011).

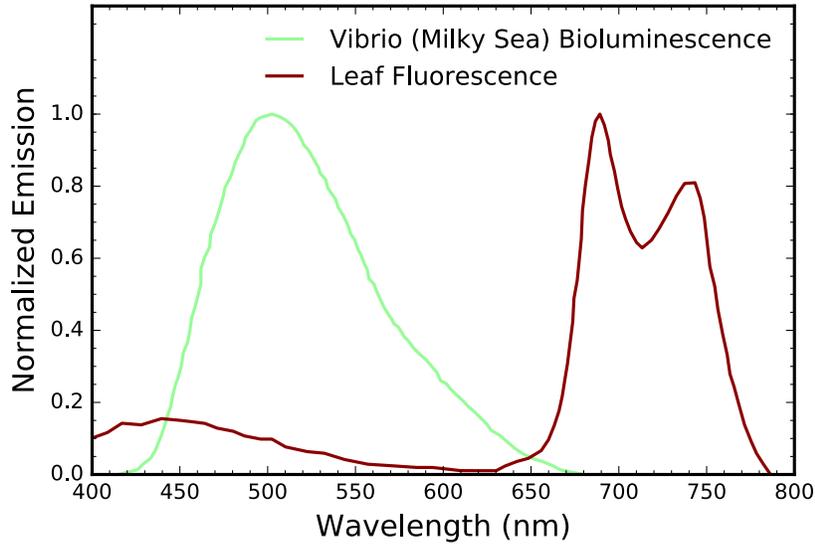

**Fig. 9** *Fluorescence and bioluminescence as biosignatures.* Bioluminescence of vibrio bacteria from Miller et al. (2005) in pale green and fluorescence of a leaf from Buschmann (2002) in dark red.

It has been proposed that photoprotective biofluourescence may be a temporal biosignature particularly suited to flaring M-stars that emit bursts of UV radiation (see O'Malley-James & Kaltenegger 2016). These authors put forward fluorescing coral proteins as a model for a hypothetical adaptation of life to high UV conditions, whereby harmful shortwave radiation from M dwarf flares is reprocessed as emitted light at longer wavelengths. The fluorescence signature would occur as light from the flare reaches the planet and would potentially have an observable spectral signature if the organisms emit light in a similar way to Earth corals and exist in sufficient abundance (O'Malley-James and Kaltenegger 2016). However, it should be noted that, as Equation 4 shows, the fluorescence signature will be a separate surface source term rather than an enhancement in the reflectivity of the surface like a red-edge signal, which would mean the intensity of the signal would not be proportional to the surface irradiation by the star at that wavelength. Overlying gaseous absorption features and auroral activity (Luger et al. 2017) will also complicate the interpretation of fluorescence biosignatures.

Fluorescence, of course, is not uniquely biological. Fluorescence is also observed in minerals such as fluorite, calcite, and many others, although their emission spectra differ markedly from those of chlorophyll and other biological proteins and could be compared to a database of known materials (e.g., O'Malley-James & Kaltenegger 2016).

Bioluminescence, unlike fluorescence, produces light *de novo* by oxidation of a light-emitting molecule generally classified as a luciferin (Haddock et al. 2010). Bioluminescence has evolved independently many dozens of times on Earth in groups as diverse as bacteria, eukaryotic marine plankton, insects, and fish (Haddock et al. 2010). Large concentrations of bioluminescent bacteria in the ocean have been observed to produce a "milky sea" effect — a faint bioluminescent glow that can span over 10,000 km$^2$ and last hours to several days (Miller et al. 2005; Fig. 9). Because bioluminescence allows the continuous production of light, it can be produced on the night side of the planet. In this case, Equation 4 is reduced to $L(\lambda) = F(\lambda)$, since the reflected component is zero. However, the only time in which an exoplanet may be observed at new phase is in the case of an "edge-on" inclination near or during transit, which may limit this technique to nearby systems. In the case of transit, bioluminescence would manifest itself with a lower apparent transit depth at the wavelengths of bioluminescent emission. However, this signal would likely need to be quite large to be detectable, possibly beyond the productivity of the biosphere. Additional study is required to further constrain this potential signature.

## *"Cryptic" Biospheres and "False Negatives" for Life*

So-called "cryptic" biospheres have been proposed that would lack surface signatures of life even if atmospheric signatures such as $O_2$ were abundant (Cockell et al. 2009). For example, types of terrestrial cyanobacteria will live under the surface layers of a rock, shielding the organism from UV radiation, but allowing photosynthetically active radiation to pass through (Cockell and Raven 2004). These cyanobacteria may be undetectable from the surface but produce $O_2$ from photosynthesis (Cockell et al. 2009). Even photosynthetic organisms that inhabit the surface may occur at too small a density to affect the disk-average of a planetary spectrum.

Alternatively, a productive biosphere with photosynthetic organisms could produce neither gaseous nor surface biosignatures (Cockell 2014). The chemosynthetic biosphere of early Earth before the advent of photosynthesis was likely to be orders of magnitude smaller than that of today (Des Marais 2000), and so, unlikely to produce observable surface or temporal signatures. Even a productive biosphere may not necessarily be remotely detectable. An emerging view from the study of Earth history suggests that during the mid-Proterozoic "boring billion" eon (1.8-0.8 Ga), both $O_2$ and $CH_4$ were undetectably low due to negative biogeochemical feedbacks (Reinhard et al. 2017), while Earth potentially lacked surface signatures of life because land plants had not yet evolved (Kenrick and Crane 1997). This lesson from Earth history warns us against interpreting the absence of evidence for life on an observably habitable planet as evidence for absence, and pushes us to adopt a cautious framework for statistical estimates of the abundance of life, given negative results.

## Conclusions

Highly productivity photosynthetic biospheres are the most likely to produce detectable surface and temporal biosignatures, though the signatures they produce need not be directly related to photosynthesis itself. The vegetation red-edge remains the most significant and well-studied surface biosignature. Detecting a direct analogue to Earth's vegetation red-edge probably requires more than a 10% surface covering fraction and a spectrophotometric sensitivity of at least ~1%. Analogue "edge" wavelengths need not be consistent with Earth chlorophyll (~700 nm), but could exist at a variety of visible and near-infrared wavelengths depending on the function of the absorbing pigment and/or the incident photon spectrum of the host star. "False positives" for biosignature 'edges' may exist in the form of band-gap generating minerals like sulfur. Linear and circular polarization could provide confirmation of pigment absorption because polarization properties are more tightly coupled to absorbing pigments than abiotic materials, though this would likely also require covering fractions of ≥10% to be detectable in the disk-averaged spectrum even with signal-to-noise ratios commensurate to observations of the Earth with interplanetary spacecraft. Temporal variations in measurable quantities such as gas concentrations, surface features, polarization properties, or surface emission of light can serve as biosignatures if they can be conceptually linked to the activities of a biosphere. For example, the seasonal variability of $CO_2$ in Earth's atmosphere fingerprints changes in the productivity of the land biosphere, though a direct-analogue (a ~2% hemisphere-averaged variation) would be challenging to detect remotely. Future work is required to constrain the range of scenarios where variability in gas concentration could detectably fingerprint an exoplanet biosphere. Fluorescence and bioluminescence are less well-studied temporal surface biosignatures. In general, analogues to surface and temporal biosignatures on Earth are relatively small effects compared to the absorption features of spectrally significant gases like $H_2O$, $CO_2$, $O_3$, and $O_2$. While longer integration times or larger telescopes may be required to detect them, surface and temporal biosignatures would provide powerful tools to confirm the presence of a biosphere on an exoplanet.

## Cross-references

- Detecting Habitability
- Atmospheric Biosignatures
- Detecting Biosignatures
- Earth's Biosignatures Through Time
- Biosignatures False Positives
- Atmospheric Evolution of a Habitable Planet

**Acknowledgements**
The author gratefully acknowledges support from the NASA Astrobiology Institute and the NASA Postdoctoral Program administered by the Universities Space Research Association. The author also appreciatively acknowledges support from and access to tools developed by the NAI Virtual Planetary Laboratory under Cooperative Agreement Number NNA13AA93A. This chapter was improved by helpful comments from Kim Bott and Victoria Meadows.

*Schwieterman, Surface and Temporal Biosignatures*